\documentclass[twocolumn,10pt,prl,aps,showpacs]{revtex4-1}
\usepackage{amsmath, amssymb, graphics}
\usepackage{graphicx,appendix}

\def\3nab{\tilde{\nabla}}

\def\hs {\,-\,}

\def\be {\begin{equation}}
\def\ee {\end{equation}}
\def\bea {\begin{eqnarray}}
\def\eea {\end{eqnarray}}

\def\case#1/#2{\textstyle\frac{#1}{#2}}
\usepackage{graphicx}
\usepackage{dcolumn}
\usepackage{bm}
\usepackage{longtable}
\def\cqg{{\em Class. Quantum Grav.\/} }
\def\grg{{\em Gen. Rel. Grav.\/} }
\def\prd{{\em Phys. Rev.\/} {\bf D}}

\def\aph{{\em Ann. Phys. (NY)\/} }
\def\plb{{\em Phys. Lett.\/} {\bf B}}

\begin{document}
\title{Cosmological dynamics of fourth order gravity: A compact view}

\author{Mohamed Abdelwahab$^{1,2}$, Rituparno Goswami$^{1,2}$ and Peter K.S. Dunsby$^{1,2,3}$}

\affiliation{1. Astrophysics, Cosmology and Gravity Centre (ACGC),  
University of Cape Town, Rondebosch, 7701, South Africa}

\affiliation{2. Department of Mathematics and Applied Mathematics,
  University of Cape Town, 7701 Rondebosch, Cape Town, South Africa}

\affiliation{3. South African Astronomical Observatory,
  Observatory 7925, Cape Town, South Africa.}

\date{\today}

\begin{abstract}
We construct a compact phase space for flat FLRW spacetimes with standard matter described by a perfect fluid with a barotropic equation of state for general $f(R)$ theories of gravity, subject to certain conditions on the function $f$. We then use this framework to study 
the behaviour of  the phase space of Universes with a non-negative Ricci scalar in 
$R+\alpha R^n$ gravity. We find a number of interesting cosmological evolutions 
which include the possibility of an initial unstable power-law inflationary point,  
followed by a curvature fluid dominated phase mimicking standard radiation, 
then passing through a standard matter (CDM) era and ultimately evolving
asymptotically towards a de-Sitter-like late-time accelerated phase.  

\end{abstract}
\pacs{98.80.Cq}
\maketitle
\section{Introduction} 
The $\Lambda$CDM (or {\em Concordance}) Model \cite{concordance} is one of the 
greatest successes of General Relativity. It reproduces beautifully all the 
main observational results e.g., the dimming of type Ia Supernovae \cite{sneIa}, 
Cosmic Microwave Background Radiation (CMBR) anisotropies \cite{cmbr}, 
Large Scale Structure formation \cite{lss}, baryon oscillations \cite{bo} and 
weak lensing \cite{wl}). Unfortunately, this model is also affected by significant 
fine-tuning problems related to the vacuum energy scale and this has led to a considerable 
amount of effort throughly exploring other viable theoretical schemes. 

Currently, one of the most popular alternatives to the {$\Lambda$CDM} model 
is based on gravitational actions which are non-linear in the Ricci curvature $R$ 
and/or contain terms involving combinations of derivatives of $R$: the so called $f(R)$ 
theories of gravity \cite{DEfR,kerner,teyssandier,magnanoff}.  
Such models first became popular in the 1980's because it was shown that they can be
derived from fundamental physical theories (Like M-theory) and naturally admit a 
phase of accelerated expansion, which could be associated with an early universe 
inflationary phase \cite{star80}.  The fact that the phenomenology of 
Dark Energy requires the presence of a similar phase (although only a late time - 
low energy one) has recently revived interest in these models. In particular, 
the idea that Dark Energy may have a geometrical origin, i.e., that there is a 
connection between Dark Energy and a non-standard behavior of gravitation on 
cosmological scales is now a very active area of research  
(see for example \cite{ccct-ijmpd,review,cct-jcap,amare,otha,perts,cct-jcap,cct-mnras}). 

Unfortunately efforts to obtain an understanding of the physics of these theories 
are hampered by the complexity of the fourth-order field equations, 
making it difficult to obtain both exact and numerical solutions, which can be 
compared with observations. Recently however, progress has been made 
in resolving these issues using a number of useful techniques. 
One such method, based on the theory of dynamical systems 
\cite{DS} has proven to be very successful in providing a simple way of 
obtaining exact solutions and a (qualitative) description of the  global dynamics of these models \cite{Dyn}. The dynamical systems analysis has up to now approached this problem in the conventional way, by first exploring the finite equilibrium points and then computing the asymptotic behavior of the phase-space using the classical method of Poincare projections. 

In this paper we develop an alternative scheme which involves compactifying the phase space for general $f(R)$ theories of gravity, subject to certain conditions on  the function $f$. We then use this framework to study the behaviour of  the phase space of Universes with a non-negative Ricci scalar in $R+\alpha R^n$ gravity. We find a number of interesting cosmological evolutions, which include the possibility, at least in principle, where the Universe begins close to at an unstable power-law inflationary equilibrium point,  then evolves towards a curvature fluid dominated phase where the effective equation of state mimics standard radiation with $w\sim 1/3$ (we will refer to such phases as radiation-like), then passes through a standard matter (CDM) era and ultimately evolves asymptotically towards a de-Sitter-like late-time accelerated phase.

We also show that as $n\rightarrow 0$, all the fixed points that approach the $\Lambda$CDM 
subspace of the complete state space of $R+\alpha R^n$ gravity,  are unstable. This implies that  the behavior of the solutions of a fourth order theory which is close to $\Lambda$CDM may be completely different from those of $\Lambda$CDM and one needs to do a careful analysis of the solutions rather than a priori assuming any global behavior of the trajectories.

\section{The field equations for fourth order gravity}

The natural extension of standard General Relativity is to consider a Lagrangian 
that contains curvature invariants of higher than linear order. In fact, renormalization 
of quantum field theory suggests that adding such terms to the standard gravitational 
action appears to be necessary \cite{Birrell} to give a first approximation to some 
quantized theory of gravity. The quadratic Lagrangians are at the first level of such modifications 
and have been studied extensively over true past two decades. The four possible second-order  curvature invariants are 
\begin{eqnarray}
R^2,\,\,\,R_{ab}R^{ab},\,\,R_{abcd}R^{abcd},\,\,\epsilon^{iklm}R_{ikst}R^{st}_{lm}\;,
\end{eqnarray} 
where $\epsilon^{iklm}$ is completely antisymmetric 4-volume element and $R$, $R_{ab}$, $R_{abcd}$  are the Ricci scalar, Ricci tensor and  Riemann tensor respectively. 
However,  for homogeneous and isotropic spacetimes, because of the following identities 
\cite{Dwitt}
\begin{eqnarray}
(\delta/\delta g_{ab})&\int & dV(R_{abcd}R^{abcd}-4R_{ab}R^{ab}+R^2)=0\;,\\
(\delta/\delta g_{ab})&\int & dV \epsilon^{iklm}R_{ikst}R^{st}_{lm}=0\;,\\
(\delta/\delta g_{ab})&\int & dV(3R_{ab}R^{ab}-R^2)=0\;,
\end{eqnarray}
it follows that the general fourth-order Lagrangian for a  these highly symmetric spacetimes contain only powers of $R$ and we can write the action as 
\be
{\cal A}= \frac12 \int d^4x\sqrt{-g}\left[f(R)+2{\cal L}_m\right]\;,
\label{action}
\ee
where ${\cal L}_m$ represents the matter contribution.

Varying the action with respect to the metric gives the following field equations:
\be
f' G_{ab} = T^{m}_{ab}+ \frac12 (f-Rf') g_{ab} 
+ \nabla_{b}\nabla_{a}f'- g_{ab}\nabla_{c}\nabla^{c}f', 
\label{field1}	
\ee
where $f'$ denotes the derivative of the function $f$ w.r.t. the Ricci scalar and  $T^{m}_{ab}$ is the matter stress energy tensor defined by
\be
T^{m}_{ab} = \mu^{m}u_{a}u_{b} + p^{m}h_{ab}+ q^{m}_{a}u_{b}+ q^{m}_{b}u_{a}+\pi^{m}_{ab}.
\ee 
Here $u^a$ is the direction of a timelike observer,  $h_{ab}$ is the projected metric on the 3-space perpendicular to $u^a$.  Also $\mu^{m}$, $p^m$, $q^{m}$ and $\pi^m_{ab}$ denotes the standard matter density,  pressure, heat flux and anisotropic stress respectively. 
Equations (\ref{field1}) reduce to the standard Einstein field equations when $f(R) = R$.

For the homogeneous and isotropic spacetimes with vanishing 3-curvature and barotropic 
perfect fluid as the standard matter source with equation of state $p=\omega\rho$, 
the independent field equations for general $f(R)$ gravity are as follows.
\begin{itemize}
\item {\em The Raychaudhuri equation}:
\begin{eqnarray}\label{Raychaudhuri}
\dot{\Theta}&=&-\dfrac{1}{3}\Theta^2-\dfrac{\rho}{2 f'}(1+3\omega)-\dfrac{f}{2f'}\nonumber\\
&& +\dfrac{R}{2}-\dfrac{\Theta}{2}\dfrac{\dot{(f')}}{f'}-\dfrac{3}{2}\dfrac{\ddot{f'}}{f'}]\;,
\end{eqnarray}
where $\Theta$ is the volume expansion of the matter flow lines $u^a$ and $\rho$ is the 
standard matter density.
\item {\em The Friedmann equation}:
\begin{eqnarray}\label{F}
\Theta^2=\dfrac{3\rho}{f'}+\dfrac{3}{2}R-\dfrac{3}{2}\dfrac{f}{f'}-3\Theta\dfrac{\dot{(f')}}{f'}\;,
\end{eqnarray}
and
\item {\em Conservation of standard matter}:
\begin{eqnarray}\label{conservation}
\dot{\rho}=-\Theta(1+\omega)\rho\;.
\end{eqnarray}
\end{itemize}
Combining the Raychaudhuri and Friedman equations, we obtain:
\begin{eqnarray}\label{combination}
R=2\dot{\Theta}+\dfrac{4}{3}\Theta^2.
\end{eqnarray}

\section{Compact phase space for positive Ricci Scalar universe:}

In this paper we will study the dynamics of Friedmann\hs Lema\^{i}tre\hs Robertson\hs Walker  (FLRW) models only in the  sector $R\geq 0$. This is because the sector $R<0$ is not of  much physical interest  and also, as we shall see later, the sectors $R>0$ and $R<0$ are connected by the invariant sub-manifold $R=0$, making the physically interesting dynamics completely confined to the sector $R>0$. Also, we consider the 3-curvature to be 
vanishing, which is an invariant sub-manifold by itself. As required by the 
{\em no\hs ghost condition} we also assume $f'>0$.
 
To compactify the phase space we rewrite the Friedmann equation (\ref{F}) in the following form:
\begin{eqnarray}\label{reF}
D^2=\dfrac{3\rho}{f'}+\dfrac{3}{2}R+\dfrac{9}{4}(\dfrac{\dot{(f')}}{f'})^2\;,
\end{eqnarray}
where
\begin{eqnarray}\label{D}
 D=\sqrt{\Big(\Theta+\dfrac{3}{2}\dfrac{\dot{(f')}}{f'}\Big)^2+\dfrac{3}{2}\dfrac{f}{f'}}\;.
\end{eqnarray}
We can now define the following set of normalized variables:
\begin{eqnarray}\label{cv}
x=\dfrac{3}{2}\dfrac{\dot{(f')}}{f'D}\qquad y=\dfrac{3}{2}\dfrac{f}{f' D^2}\qquad 
\Omega_m=\dfrac{3\rho}{f' D^2}\nonumber\\
z=\dfrac{3}{2} \dfrac{R}{D^2}\qquad Q=\dfrac{\Theta}{D}\;. 
\end{eqnarray}
To guarantee that the propagation equations for these compact variables will result in a 
dimensionless dynamical system, we need to define a new time variable $\tau$, such that
\begin{eqnarray}
\dfrac{d}{d\tau}\equiv '=\dfrac{1}{D}\dfrac{d}{dt}\;.
\end{eqnarray}
For $\tau$ to be a monotonously increasing time variable, a normalization D  is chosen such that it is strictly positive at all times. It is clear by  construction that when $\Theta=0$ the normalized dynamical variables as well as the  time variable are well defined, thus this normalization allows the study of general static, re-collapsing and bouncing solutions. 

From the Friedmann equation we obtain the following constraints,
\begin{eqnarray}\label{const}
\Omega_m+z+x^2 &=&1\;,\nonumber\\
(Q+x)^2+y&=&1.
\end{eqnarray}
The first constraint comes directly from Friedmann equation,  while the second one arises from the definition of the normalization parameter D. 
According to these constraints and considering $R>0$, $\rho>0$ and $f'>0$, 
we see that the above dynamical variables have to be defined in the 
following ranges, 
\begin{eqnarray}\label{range}
0\leq\Omega_m \leq 1\;,\qquad 0\leq z \leq 1\;,\qquad -1\leq x \leq 1\nonumber\\
-2\leq Q\leq 2; \qquad 0\leq y\leq 1\;,
\end{eqnarray}
making the complete phase space compact. Also since the variable $Q$ is a normalized 
Hubble parameter, the cosmological solutions will naturally include both expanding and collapsing as well as static solutions and  these two sets of solutions are connected via the non-invariant subset $Q=0$. 

\section{The propagation equations}

An autonomous system, which is equivalent to cosmological equations 
(\ref{Raychaudhuri}-\ref{combination}) can be derived by differentiating the compact variables 
(\ref{cv}), with respect to $\tau$ and using (\ref{Raychaudhuri}-\ref{conservation}).
The dimensionality of the resultant system can then be reduced by using the two constraints (\ref{const}). By eliminating the dynamical variables $\Omega_m$ and $y$, we obtain the following 3-dimensional effective autonomous system:
\begin{widetext}
\begin{eqnarray}\label{DS}
x'&=&\dfrac{1}{6}\Big(-3(1+\omega)-(1+3\omega)x^4-4Q^2(-1+x^2)+(1+3\omega)z\nonumber\\
&-&Qx\Big[(5+3\omega)(-1+x^2)+3(1+\omega)z \Big]+
x^2\Big[4+6\omega+z(-3(1+\omega)-2\Gamma) \Big] \Big) \nonumber\\
z'&=&\dfrac{z}{3}\Big(-4Q^2x-Q((5+3\omega)x^2+3(1+\omega)(-1+z))\nonumber\\
&+&x(5+3\omega-(1+3\omega)x^2-\Gamma\Big[-2-3n(1+\omega)+2z+3n(1+\omega)(z+(Q+x)^2)\Big]\Big)\\
Q'&=&\dfrac{1}{6}\Big(-4Q^2x-Q\Big[(5+3\omega)x^2+3(1+\omega)(-1+z)\Big]\nonumber\\
&+&x(5+3\omega-(1+3\omega)x^2-\Gamma\Big[-2-3n(1+\omega)+2z+3n(1+\omega)((Q+x)^2+z)\Big]\Big)\;\nonumber,
\end{eqnarray}
\end{widetext}
where $\Gamma\equiv {f'}/{R f''}$. In general, the system is not closed unless $\Gamma$ 
is expressed in terms of the dynamical variables (\ref{cv}). For example, in the case 
of $R+\alpha R^n$, we have
\begin{eqnarray}
\Gamma \equiv -\dfrac{z}{n(y-z)}=\dfrac{z}{n[(q+x)^2+z-1]}\,.
\label{gamma}
\end{eqnarray}
Thus, the above system defines the dynamics of all well defined $f(R)$ theories for which 
${f'}/{R f''}$ is invertible in terms of the dynamical variables. From equations (\ref{DS}) we can see that $z=0$ is an invariant sub-manifold. and in the $z=0$ 2-surface the line $Q=0$ is an invariant subset.
Since $z=0$ corresponds to $R=0$, we obtain an important result:

{\em For all well defined functions $f(R)$, with $f'>0$ and ${f'}/{R f''}$ invertible 
in terms of the dynamical variables defined by (\ref{cv}), a FLRW universes with non\hs negative Ricci Scalar, continues to be so, both in the future and in the past. Also an $R=0$ universe  can never undergo a bounce in the future or the past.}

In the next section we will fix the function $f$ to be the class of theories $f(R)=R+\alpha R^n$ and study the dynamics of the flat FLRW universes and their stability for those theories. In order to study the stability of the fixed points of the dynamical systems  (\ref{DS}), we will use the very well known techniques, which involve linearizing the dynamical equations around the equilibrium points and then finding the eigenvalues of the linearization matrix (the Jacobian) at the equilibrium points. If the Jacobian is well defined, then they can be classified according to the sign of the real part of eigenvalues as attractors, repellers and saddle points. 

\section{The fixed points and Exact solutions for  $R+\alpha R^n$ gravity}

As we have seen from the equation (\ref{gamma}), ${f'}/{R f''}$ is invertible 
in terms of the dynamical variables for $f(R)=R+\alpha R^n$. It is interesting to note 
that the constant `$n$' couples to the dynamical equations (\ref{DS}) only via the 
quantity $\Gamma$ and the constant $\alpha$ does not couple to the equations at all. 
Hence all the fixed point of the system are necessarily independent of $\alpha$.    

The coordinates of the fixed points are shown in Table 1. Note that each fixed point 
has an expanding ($Q>0$) and a collapsing ($Q<0$) version as indicated by the subscripts ($+$; $-$) respectively. Also some points only occur in the compact state space defined by (\ref{range}) for certain ranges of $n$. The occurrence of the fixed points outside the compact region for specific $n$ and $\omega$ means that the constraints (\ref{const}) are not satisfied and consequently these fixed points are not physical for these values of $n$ and $\omega$). Fixed points that are not physical for these values of $n$ and $\omega$ have been excluded from the analysis. 
\begin{widetext}
\textbf{Table 1}. 
Coordinates of the equilibrium points for $R+\alpha R^{n}$-gravity. We will not explicitly state the expressions for s,g1,...,g4 and f1,...,f4, which are rational functions of n and $\omega$, however we give them at the following link \cite{link}.
\begin{center}
\tabcolsep 5.8pt
    \small
\begin{tabular}{l|l|l}
\hline
Fixed points  & Coordinates $(x,\Omega,z,Q)$  & Solution $a(t)$ \\
\hline
\hline
$A_\pm$     & $(1,0,0,\pm 2)$ & $a_0\sqrt{t-t_0}$\\
$B$ & $(\pm 1,0,0,0)$ &$a_0$\\
$C$ &$(-\dfrac{\sqrt{3+12\omega+9\omega^2}}{1+3\omega},-\dfrac{2}{1+3\omega},0,0)$  &$a_0$\\
$D_\pm$ & $(\dfrac{1-3\omega}{3(\omega-1)},-\dfrac{4(3\omega-2)}{9(\omega-1)^2},0,\pm \dfrac{2}{3(\omega-1)})$ &$a_0 \sqrt{t-t_0}$\\
$E_\pm$ & $(0,0,1,\pm\dfrac{1}{\sqrt{2}})$ &$a_0 e^{C t}$\\
$F_\pm$ &$(f_1(n,\omega),g_1(n,\omega),l_1(n,\omega),n_1(n,\omega))$ &$a_0 \sqrt{t-t_0}$\\
$G_\pm$ & $(f_2(n,\omega),g_2(n,\omega),l_2(n,\omega),n_2(n,\omega))$ &$a_0 (t-t_0)^{s(n,\omega)}$\\
$I_\pm$ &$(f_3(n),g_3(n),l_3(n),n_3(n))$  &$a_0 ((n-2)t-t_0)^{\dfrac{-1+3n-2n^2}{-2+n}}$ \\
$L_\pm$ &$(f_4(n,\omega),g_4(n,\omega),l_4(n,\omega),n_4(n,\omega))$  &$a_0 (3t(1+\omega)-t_0)^{\dfrac{2n}{3(1+\omega)}}$\\
$N_\pm$ &$(0,\dfrac{2}{3},\dfrac{1}{3},\pm\dfrac{\sqrt{6}}{3})$ & $a_0(2 t-t_0)^{2/3}$\\
\hline
\hline
\end{tabular}
\end{center}
\end{widetext}
By looking at the coordinates of the fixed points in Table 1, we can distinguish 
two classes;  the first corresponds to points with coordinates that are 
independent of $n$, which means that these points are common to all $f(R)$ theories. 
This class contains the fixed points $A_\pm,B,C_\pm,D_\pm,E_\pm$ and $N_\pm$ and they all lie on the boundary of the compact region except for the point $N$. 

In the non\hs compact analysis developed in \cite{Sante 1}, non of these boundary points appear. Furthermore, even though $N_\pm$ is not a boundary point, it doesn't appear in \cite{Sante 1}, because of it's special location in the phase space \hs  it lies exactly on the intersection of the plane  $x=0$ and the surface $z=y=1-(Q+x)^2$. In this case one has to take the limit of $\Gamma$ carefully as one approaches this point and the standard techniques of finding fixed points breaks down for this case. 

The other class contains fixed points with coordinates that depend on n and  $\omega$;  this class contains the three points $L_\pm,I_\pm$ and $F_\pm$.  $F_\pm$ is the only boundary point and it lies in the invariant sub-manifold $z=0$. The expanding versions of the points $L_\pm$ and $I_\pm$ correspond to the equally labeled finite points in \cite{Sante 1}. The point H in \cite{Sante 1} enters the compact sector, which we consider in this paper only when $n=(1+\sqrt{3})/2$ and for this value of $n$ it merges with the point $I$. All the other points that appear in the above mentioned reference do not appear in the sector we are studying in this paper.\newpage
\begin{widetext}
\textbf{Table 2}. The stability of the fixed points for $\omega=0;1/3$.\\
\begin{tabular}{l|l|l|l|l}
  \hline
 Fixed point &\multicolumn{2}{|c|}{Physical range} & \multicolumn{2}{|c|}{Stability}\\
  \hline
  Equation of state&$\omega=0$ &$\omega=\dfrac{1}{3}$&$\omega=0$&$\omega=\dfrac{1}{3}$ \\
    \hline
    \hline
 $A_-$ &$\forall$n & $\forall$n &Attractor &Attractor \\
 $A_+$ &$\forall$n & $\forall$n &Repeller&Repeller \\
 $B$  &$\forall$n & $\forall$n &Attractor &Attractor \\
 $D_\pm$ &$\forall$n& $\forall$n &Saddle&Saddle\\
 $E_-$ &$\forall$n & $\forall$n &Repeller&Repeller \\
 $E_+$&$\forall$n & $\forall$n &Attractor for $n\in (0,2)$ &Attractor for $n\in (0,2)$ \\
 $F_\pm$ &$n\in (0,\dfrac{1}{3}+ \dfrac{\sqrt{57}}{9})$&$n\in (0,\dfrac{1}{8}+\dfrac{\sqrt{17}}{8})$&Saddle&Saddle\\
$I_-$ &$n\in (\dfrac{1}{2},1)$ and $n >5/4$&$n\in (\dfrac{1}{2},1)$ and $n >5/4$& Saddle for $n\in (1/2,1)$& Saddle for $n\in (1/2,1)$ \\
&&&Attractor for $n=5/4$&Attractor for $n=5/4$\\
&&&Saddle for $n\in (5/4,2)$&Saddle for $n\in (5/4,2)$\\
&&&Attractor for $n>2$&Attractor for $n>2$ \\
$I_+$ &$n\in (\dfrac{1}{2},1)$ and $n >5/4$&$n\in (\dfrac{1}{2},1)$ and $n >5/4$& Saddle for $n\in (1/2,1)$& Saddle for $n\in (1/2,1)$ \\
&&&Repeller for $n=5/4$&Repeller for $n=5/4$\\
&&&Saddle for $n\in (5/4,2)$&Saddle for $n\in(5/4,2)$\\
&&&Repeller for $n>2$&Repeller for $n>2$ \\
$L_\pm$ &$n \in (\dfrac{3}{4},\dfrac{4}{7}+\dfrac{\sqrt{37}}{7})$&$n \in (1,\sqrt{2})$&Saddle&Saddle \\   
$N_-$ &$\forall$n&$\forall$n&$\text{Spiral}^+$ &$\text{Spiral}^+$ \\   
$N_+$ &$\forall$n&$\forall$n&$\text{Spiral}^-$ &$\text{ Spiral}^-$ \\   
\hline
\end{tabular}
\end{widetext}

\subsection{Exact solutions}

The exact solutions at the fixed points are also summarized in Table 1 and the stability analysis for the dust and radiation cases are summarized in Table 2. First we discuss the static solutions. From definition (\ref{DS}), $Q=0 \Rightarrow \Theta=0$, so any fixed point that lies on the surface $Q=0$ represents a static Universe. By looking at the coordinates of the fixed points in Table 1,  we can see that the point $B$ is static for all values of n and $\omega$.  The point $I_\pm$ is static only for $n=1/2$ and $n=1$ and we find that for these values of $n$ the point $I_\pm$ represent an unstable saddle point.

We now proceed to find the exact solutions for the scale factor at the non\hs static fixed points. The expansion rate and the deceleration parameter $q=-\dfrac{\ddot{a} a}{\dot{a}^2}$ are related by the Raychaudhuri equation,
\begin{eqnarray}\label{deceleraion}
\dot{\Theta}=-\dfrac{1}{3}(1+q) \Theta^2\;.
\end{eqnarray}
If we know the value of the deceleration parameter $q_i$ at some fixed point i, 
we can use the above equation to obtain the behavior of the scale factor at that point. When $q_i =-1$ we have de\hs Sitter solutions $(\Theta=constant)$ or static solutions $(\Theta =0)$. For $q_i =0$ we have a Milne evolution and when $-1<q_0 <0$ ; $q_0 >0$ we have accelerated and decelerated power law behaviours respectively. 

To obtain the exact solutions for the scale factor $a(t)$ associated with the non\hs static $\Theta\neq 0$ equilibrium points we need to have an expression for $q$ in term of the compact variables. From the definition $q$ we obtain:  
\begin{eqnarray}\label{q in term of Compact v}
q_i=1-\dfrac{z_i}{Q_i^2}\;.
\end{eqnarray}
The non-invariant surface $z_i=Q_i^2$ is the transition surface between accelerated and decelerated expansions phases (see Fig 1). By substituting \ref{q in term of Compact v} 
in \ref{deceleraion} we obtain:
\begin{eqnarray}\label{scale factor}
\dot{\Theta}=-\dfrac{1}{3} (2-\dfrac{z_i}{Q_i^2})\Theta^2\;,
\end{eqnarray}
where $Q\neq 0$. The evolution of the scale factor can now be given 
directly by integrating equation \ref{scale factor}:
\begin{eqnarray}
a(t)=a_0(t-t_0)^{\beta_i},
\end{eqnarray}
where
\begin{eqnarray} 
\beta_i=(2-\dfrac{z_i}{Q_i^2})\;.
\end{eqnarray}
The constants of integration can be obtained by substituting the solutions into the original equations. As explained in \cite{Sante 1}, these solutions must satisfy all the cosmological equations in order to be considered physical.

By looking at Table 1 we can distinguish two classes of non\hs static solutions. The first class $\{A_\pm,D_\pm,E_\pm,F_\pm$ and $N_\pm$\} contain solutions that are independent of n and $\omega$. The fixed point B represents a static phase as mentioned earlier, the points A, D and F are radiation\hs like phases.  The expanding version of the point A is a saddle for $z>0$ and repeller for $z=0$ and the other two points D an F are saddles. The fixed point N represent a matter phase and the expanding version of this point is a spiral$^-$.

The evolution of the scale factor a(t) for the fixed point $L$ is a function of $(n,\omega,t)$ and for fixed point $I$ is function of $(n,t)$. The dependence of these solutions on n and/or $\omega$ provide us with additional degrees of freedom that can lead to interesting cosmological scenarios.

When $\omega=0$, the fixed point $L$ merges with $N$ for $n=1$, and it merges with the point $D_\pm$ for $n=3/4$. When $\omega=1/3$ it merges with $D_\pm$ for $n=1$, and for $n=4/3$ it corresponds to the matter point $N$. In the case $\omega=0$ or $\omega=1/3$, we find that for all values n for which this point is physical, the expansion (contraction) is never accelerating.

As mentioned earlier, the evolution of the scale factor for the point $I$ is independent of the equation of state parameter $\omega$. For $n=5/4$ the 
fixed point $I$ merges with $A$, for $n=2$ it merges with point $E$ 
and for $n=7/12+\sqrt{73}/12$ it is a matter point. 

We also find that for this fixed point the expansion (contraction) is accelerating for $n>1/2(1+\sqrt{3})$. The existence of this accelerated phase together with the fact that for $n>1/2(1+\sqrt{3})$ the point $L$ is a matter-like point,  leads to the possibility of an extremely interesting cosmological scenario, where it is possible in principle to find an orbit that starts close to the unstable accelerating phase $I_+$, evolves past the unstable radiation\hs like point $D_+$, followed by the unstable matter point $L_+$ and finally ends up at the de\hs Sitter attractor $E_+$. 

In Figs. 3 and 4 we have plotted two interesting orbits. The orbit in Fig 3 is for $\omega=0$ and $n=5/4$. It begins near the radiation\hs like points $A_+/I_+$,  passes nearby the radiation\hs like point $D_+$, followed by the standard matter point $L_+$ and ends up at the de-Sitter attractor $E_+$.  The orbit in Fig 4 is for $\omega=-11/18+\sqrt{73}/18$ and $n=7/12+\sqrt{73}/12$. It begins near the radiation\hs like point $A_+$, passes nearby the matter point $I_+$, then close to the matter point $L_+$ and ends up at the de Sitter attractor $E_+$.

It is interesting to see that for all fixed points whose  x\hs coordinate goes to zero as $n\rightarrow 0$, are unstable. As 
the $\Lambda$CDM subspace lies on the $x=0$ surface and this surface is not an invariant sub\hs manifold, this implies that 
the behavior of the solutions of a fourth order theory which is close to $\Lambda$CDM, may be completely different from those of $\Lambda$CDM. Furthermore,  this suggests that the best  fit model to the current observational data within the complete state space of $R+\alpha R^n$,  may be given by a non-infinitesimal value of $n$.  

\begin{figure}[htb!]
\begin{center}
\includegraphics[width=8cm]{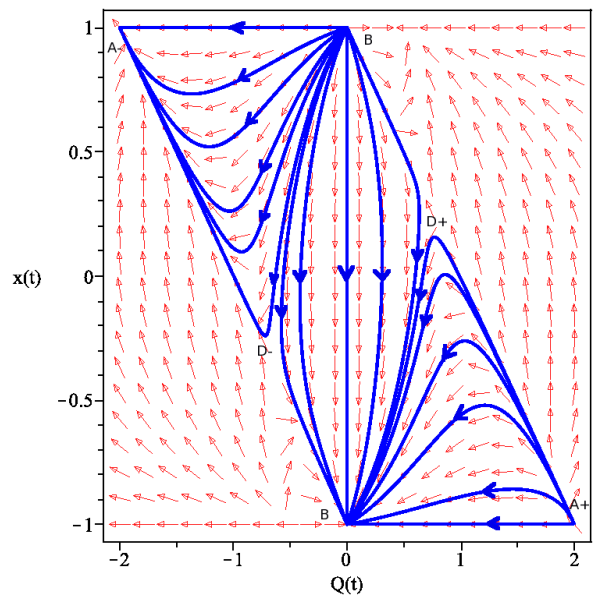}
\end{center}
\caption{plot of the invariant subspace  $z=0$ for $\omega =0$ ;$n={5}/{4}$. The left half of the state space corresponds to collapsing  models, while the right half contain expanding models. This is indicated by the subscripts of the various equilibrium points.}
\end{figure}

\begin{figure}[htb!]
\begin{center}
\includegraphics[width=9cm]{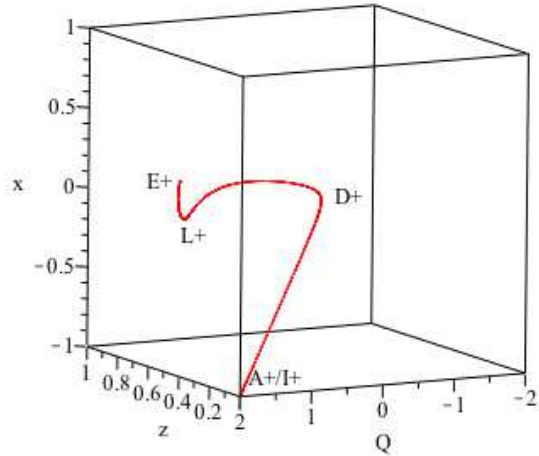}
\end{center}
\caption{For this orbit $\omega=0$ and $n=5/4$.}
\end{figure}

\begin{figure}[htb!]
\begin{center}
\includegraphics[width=9cm]{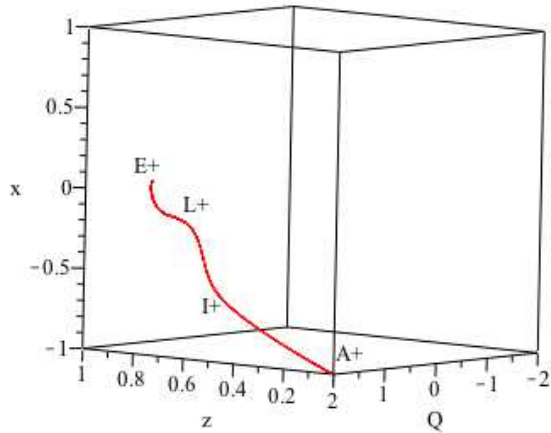}
\end{center}
\caption{For this orbit $\omega=-11/18+\sqrt{73}/18$ and $n=7/12+\sqrt{73}/12$.}
\end{figure}

\section{Conclusion}
In this work, we have presented a careful analysis of the state space of the class of $R+\alpha R^n$ theorys of gravity,  focusing on the $R>0$ sector with $K=0$, together with the no-ghost condition $f(R);f'(R)>0$.

Due to the complexity of this class of gravity theories, the standard Hubble normalization does not lead to a compact dynamical variables. In order to construct variables defining a compact dynamical system one has to use an appropriate normalization. In this paper we used the same formalism used in \cite{Naureen}, where we absorbed all the negative contributions of the Friedmann equation into the normalization.  First of all we obtained the following important result: For all well defined function $f(R)$, with $f'>0$ and ${f'}/{R f''}$ invertible in terms of the dynamical variables defined by $(14)$, the FLRW universes with non-negative Ricci Scalar, continues to be so both in the future and in the past. Also an R = 0 universe can never undergo a bounce in the future or past.

Our compact analysis shows that there are more equilibrium points than in the corresponding non-compact analysis in \cite{Sante 1}. In particular we find a new finite fixed point $N_\pm$. Because of it's very special location in the phase space, it is quite difficult to obtain this point using the standard techniques. This point is found to represent a matter phase and the expanding version of this point is $spiral^{-}$.

Furthermore, we find that for $n> 1/2(1+\sqrt{3})$ the phase space of $R+\alpha R^n$, contains two accelerated fixed points $E_+;I_+$, together with two other saddle points (one represent a radiation phase $D_+$ and the other represent a matter-like phase $L_+$). Although we have obtained all the desired fixed points and desired stability, this does not necessarily imply that there is an orbit connecting them. Due to the fact that for $n> 1/2(1+\sqrt{3})$, the two accelerated points and the matter-like point are quite close to each other in the phase space, which makes it difficult to prove the existence of an orbit connecting these points together with the radiation\hs like point. But the presence of all these phases in the state space of $R+\alpha R^n$ makes a more detailed investigation worth pursuing.


\begin{thebibliography}{99}
\bibitem{concordance}
J. P. Ostriker and P. J. Steinhardt, Cosmic Concordance,  [arXiv:astro-ph/9505066].

\bibitem{sneIa}S. Perlmutter {\it et al.}, Astrophys. J. {\bf 517}, 565 (1999); A. G. Riess {\it et al.}, Astron. J. {\bf 116}, 1009 (1998);  J. L. Tonry {\it et al.},
{\em Astrophys. J.}  {\bf 594}, 1 (2003); R. A. Knop {\it et al.},
{\em Astrophys. J.} {\bf 598}, 102 (2003); A. G. Riess {\it et al.}
{\em Astrophys. J.} {\bf 607}, 665 (2004); S. Perlmutter {\it et al.}
{\em Astrophys. J.} {\bf 517}, 565 (1999); {\em Astron. Astrophys.} {\bf 447},
31 (2006).

\bibitem{cmbr} D. N. Spergel {\it et al.} {\em Astrophys. J. Suppl.} {\bf 148}, 175 (2003); 
D. N. Spergel {\it et al.}, {\em Astrophys .J. Suppl.} {\bf 170}, 377 (2007). 

\bibitem{lss}M. Tegmark {\it et al.}, Phys. Rev. {\bf D 69}, 103501 (2004); U. Seljak {\it et al.}, Phys. Rev. {\bf D 71}, 103515 (2005); S. Cole {\it et al.}, Mon. Not. Roy. Astron. Soc. {\bf 362}, 505 (2005).

\bibitem{bo} D. J. Eisenstein {\it et al.}, Astrophys. J.
{\bf 633}, 560 (2005); C. Blake, D. Parkinson, B. Bassett, K.
Glazebrook, M. Kunz and R. C. Nichol, {\em Mon. Not. Roy. Astron. Soc.}
{\bf 365}, 255 (2006).

\bibitem{wl} B. Jain, A. Taylor,  {\em Phys. Rev. Lett.} {\bf 91}, 141302 (2003).

\bibitem{Dunajski:2008tg}
 M.~Dunajski and G.~Gibbons,
  Class.\ Quant.\ Grav.\  {\bf 25}, 235012 (2008)
 [arXiv:0807.0207 [gr-qc]].

\bibitem{star80}  A. A. Starobinsky, {\em Phys. Lett.} {\bf B 91}, 99 (1980); K.~S.~Stelle,
{\em Gen. Rel. Grav.} {\bf 9} 353 (1978).

\bibitem{DEfR} S. M. Carroll, V. Duvvuri, M. Trodden and M. S. Turner
\prd {\bf 70} 043528 (2004); S. Nojiri and S. D. Odintsov \prd {\bf 68}, 123512 (2003); S. Capozziello, {\it Int. Journ. Mod. Phys.} D {\bf 11}, 483 (2002); V. Faraoni \prd {\bf 72}, 124005 (2005); M. L. Ruggiero and L. Iorio  JCAP {\bf 0701} 010 (2007); A. de la Cruz-Dombriz and A. Dobado A, \prd {\bf 74}, 087501 (2006); N. J. Poplawski,  \prd {\bf 74}, 084032 (2006); N. J. Poplawski, \cqg {\bf 24}, 3013 (2007);
A. W. Brookfield, C. van de Bruck and L. M. H. Hall, \prd {\bf 74}, 064028 (2006); Y. Song, W. Hu and I. Sawicki \prd {\bf 75}, 044004 (2007); B. Li, K. Chan and M. Chu, \prd {\bf 76}, 024002 (2007); X. Jin, D. Liu and X. Li. [arXiv: astro-ph/0610854]; T. P. Sotiriou and S. Liberati S \aph {\bf 322}, 935 (2007); T. P. Sotiriou, \cqg {\bf 23}, 5117 (2006); R. Bean, D. Bernat, L. Pogosian, A. Silvestri and M. Trodden \prd {\bf 75}, 064020 (2007); I. Navarro and K. Van Acoleyen, JCAP {\bf 0702}, 022 (2007); A. J. Bustelo and D. E. Barraco \cqg {\bf 24}, 2333 (2007); G. J. Olmo \prd {\bf 75}, 023511 (2007); J. Ford, S. Giusto and A. Saxena, {\em Nucl. Phys.} B {\bf 790}, 258 (2008); F.  Briscese, E. Elizalde, S. Nojiri and S. D. Odintsov, \plb {\bf 646}, 105 (2007); S. Baghram,  M. Farhang and S. Rahvar,  \prd {\bf 75}, 044024 (2007); D. Bazeia, B. Carneiro da Cunha,  R. Menezes and A. Petrov \plb {\bf 649}, 445 (2007); P. Zhang \prd {\bf 76}, 024007 (2007); B. Li and J. D. Barrow 
\prd {bf 75}, 084010 (2007);  T. Rador,  {\em Phys. Lett.} B {\bf 652}, 228 (2007); T. Rador, \prd {\bf 75}, 064033 (2007); L. M. Sokolowski,  {\em Class. Quant. Grav.} {\bf 24}, 3391 (2007); V. Faraoni, \prd {\bf 75}, 067302 (2007); O. Bertolami, C. G. Boehmer, T. Harko and F. S. N. Lobo F S, \prd {\bf 75}, 104016 (2007); S. K. Srivastava, {\em Int. J. Theor. Phys.} {\bf 47}, 1966 (2008); S. Capozziello, V. F. Cardone and A. Troisi, JCAP {\bf 08}, 001 (2006); A. A. Starobinsky, {\em JETP Lett.} {\em 86}, 157 (2007); A. de Felice, S. Tsujikawa, f(R) theories, arXiv:1002.4928v1; E. Elizalde, S. D. Odintsov, L. Sebastiani, S. Zerbini, 	arXiv:1108.6184v1 [gr-qc]; E. Elizalde, S. Nojiri, S.D. Odintsov, L. Sebastiani, S. Zerbini, {\em Phys.Rev.} D {\bf 83} 086006 (2011);S. Nojiri, S.D. Odintsov {\em Phys.Rept.} {\bf 505} 59-144, (2011); S. Nojiri, S.D. Odintsov  {\em Int.J.Geom.Meth.Mod.Phys.} {\bf 4} 115-146, (2007).

\bibitem{kerner}  R. Kerner \grg {\bf 14}, 453 (1982);
J. P. Duruisseau, R. Kerner, \cqg {\bf 3},  817 (1986).

\bibitem{teyssandier} P. Teyssandier, \cqg {\bf 6} 219 (1989).

\bibitem{magnanoff} G. Magnano, M. Ferraris and M. Francaviglia, \grg {\bf 19} 465 (1987).

\bibitem{ccct-ijmpd} S. Capozziello, V. F. Cardone, S. Carloni,
A. Troisi, {\em Int. J. Mod. Phys.}  D {\bf 12}, 1969 (2003).

\bibitem{review} S. Capozziello, S. Carloni and A. Troisi, {\em Recent Res. Devel. Astronomy \& Astrophysics} { \bf 1}, 625 (2003).

\bibitem{cct-jcap} S. Capozziello, V. F. Cardone, A. Troisi, {\em JCAP} {\bf 0608}, 001 (2006).
\bibitem{amare} Amare Abebe, Mohamed Abdelwahab, Alvaro de la Cruz-Dombriz, Peter K.S. Dunsby, [arXiv:1110.1191].
\bibitem{otha}K.~i.~Maeda and N.~Ohta,
 {\em Phys.\ Lett.}\  B {\bf 597}, 400 (2004);  K.~i.~Maeda and N.~Ohta,
 {\em Phys.\ Rev.}\  D {\bf 71}, 063520 (2005); N.~Ohta, {\em Int. J. Mod. Phys.} A {\bf 20}, 1 (2005);  K.~Akune, K.~i.~Maeda and N.~Ohta, {\em Phys. Rev.}  D {\bf 73}, 103506 (2006).

 \bibitem{perts}
 S. Carloni, P. K. S. Dunsby, A. Troisi, {\em Phys. Rev.} D {\bf 77} 024024 (2008); 
 K. N. Ananda, S. Carloni, P. K. S. Dunsby, {\em Phys. Rev.} D {\bf 77}, 024033 (2008);
 K. N. Ananda, S. Carloni, P. K. S. Dunsby, {\em Class. Quant. Grav.} {\bf 26} 235018 (2009);
K. N. Ananda, S. Carloni, P. K. S. Dunsby, A characteristic signature of fourth order gravity, [arXiv:0812.2028].

\bibitem{cct-mnras} S. Capozziello, V.F. Cardone, A. Troisi, 
{\em Mon.  Not.  Roy.  Astron.  Soc.} {\bf 375}, 1423 (2007).

\bibitem{DS}
J. Wainwright and G. F. R. Ellis (ed.), 
Dynamical Systems in Cosmology. Cambridge: Cambridge University Press. ISBN 0-521-55457-8.

\bibitem{link} 
http://www.mth.uct.ac.za/$\sim$peter/coordinates.pdf

\bibitem{Dyn}
S. Carloni, P.  K. S. Dunsby, S. Capozziello, A. Troisi, {\em Class. Quant. Grav.} {\bf 22}, 4839 (2005); S. Carloni and P. K. S. Dunsby,  {\em J. Phys.}  A {\bf 40}, 6919 (2007); S. Carloni, A. Troisi, P. K. S. Dunsby, {\em Gen. Rel. Grav.} {\bf 41}, 1757 (2009).

\bibitem{Dwitt}
B. de Witt, Dynamical Theory of Groups and Fields
(New York: Gordon and Breach, 1965)
\bibitem{Birrell}
Birrell N, Davies P 1982 Quantum Fields in Curved Space (Cambridge: Cambridge Univ. Press)
\bibitem{sexl}
Pechlaner E and Sex1 R 1966 Commun. Math. Phys. 2 165
\bibitem{Sante 1}
S. Carloni, K. N. Ananda, P. K. S. Dunsby and M. E. S. Abdelwahab, arXiv:0812.2211 [astro-ph].
\bibitem{Naureen1} N. Goheer, J. A. Leach and P. K. S. Dunsby, {\em Class. Quant. Grav.} {\bf 25},  035013 (2008).
\bibitem{Naureen} N. Goheer, R. Goswami, and P. K. S. Dunsby, {\em Class. Quant. Grav.} {\bf 26},  105003 (2009).
\end{thebibliography}
\end{document}